\documentclass[APS,twocolumn,showpacs,preprintnumbers,amsmath,amssymb]{revtex4}

\usepackage{graphicx}
\usepackage{amsfonts,amssymb, amsmath}
\usepackage{myPackage}
\usepackage{color}
\usepackage{psfrag}

\newcommand{\myItem}{\vspace{.2em}\\$\bullet$~}
\newcommand{\myItemEnd}{\vspace{.2em}\\}

\begin{document}

\title{One-shot entanglement generation over long distances in noisy quantum networks}
\author{S. Perseguers$^1$}
\author{L. Jiang$^2$}
\author{N. Schuch$^1$}
\author{F. Verstraete$^3$}
\author{M.D. Lukin$^2$}
\author{J.I. Cirac$^1$}
\author{K.G.H. Vollbrecht$^1$}
\affiliation{
    $^1$Max-Planck--Institut f\"ur Quantenoptik, Hans-Kopfermann-Strasse 1, D-85748 Garching, Germany\\
    $^2$Physics Department, Harvard University, MA-02138 Cambridge, USA\\
    $^3$Fakult\"at f\"ur Physik, Universit\"at Wien, Boltzmanngasse 5, A-1090 Wien, Austria
}
\date{\today}

\begin{abstract}
We consider the problem of creating a long-distance entangled state between two
stations of a network, where neighboring nodes are connected by noisy quantum
channels. We show that any two stations can share an entangled pair if the effective
probability for the quantum errors is below a certain threshold, which is achieved
by using local redundant encoding to preserve the global phase and network-based
correction for the bit-flip errors. In contrast to the convensional quantum repeater
schemes we are not limited by the memory coherence time, because all quantum
operations only use one-way classical communication and can be done in one shot.
Meanwhile, the overhead of local resources only increases logarithmically with the
size of the network, making our proposal favorable to practical applications of
long-distance quantum communication.
\end{abstract}

\maketitle

\section{Introduction}
\label{sec:intro}
The task of creating entanglement over long distances has attracted a lot of
attention in the last ten years and has been shown to be feasible using the so-called
quantum repeaters protocols \cite{BDCZ98,DBCZ99,CTSL05}, which intersperse connection
and purification steps. One of the remaining problem for their technical realization
is the development of efficient and reliable quantum memories \cite{HKBD07}. An
alternative to this (one-dimensional) repeaters method is to exploit the high
connectivity of quantum networks with percolation strategies \cite{ACL07,PCA+08}.
However, these last results hold only for pure states and perfect local quantum
operations. We propose here to combine the idea of quantum error correcting
codes for fault-tolerant communication \cite{KL96} and some features of two-dimensional
quantum networks. This allows us to separately treat the bit-flip and the phase errors
that occur when imperfect quantum operations are made on the qubits. Compared
with one-dimensional quantum repeaters with encoding \cite{JL08}, which are based on general
quantum error correction and thus quite resource-consuming, our network can use smaller
and more efficient quantum codes that correct one specific type of error (the phase
error), while the bit-flip errors are corrected by the network.

%
\subsection*{Description of the model}
\label{ssec:intro-desc}
\begin{figure}
    \begin{center}
        \includegraphics[height=2.5cm]{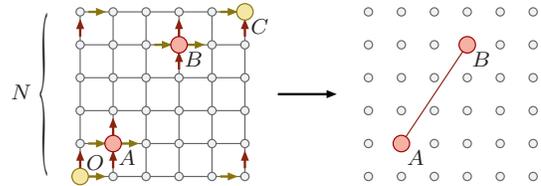}
        \caption{A quantum network with a station located at each vertex of a
            $N\times N$ square lattice. The goal is to create a long-distance entangled
            state between some previously chosen stations $A$ and $B$.}
        \label{fig:model}
    \end{center}
\end{figure}

The quantum network we consider throughout the paper consists in a square lattice
where nodes represent the stations and edges the quantum channels, see Fig.~\ref{fig:model}.
Quantum states can be transmitted through these channels, and entanglement (i.e.,
short-distance Bell states) can be created between neighboring stations.
We would like to use these local resources to generate long-distance Bell pairs
$\ket{\Phi^+}_{AB}=\ket{00}_{AB}+\ket{11}_{AB}$ between some chosen destination stations $A$ and $B$.
We assume perfect classical communication among all stations, but imperfect quantum
operations and channels, so that the local entangled pairs have limited fidelity
(more details on the error model is given in App.~\ref{app:error}).
%
\section{Network with bit-flip errors only}
\label{sec:flip} We assume for the moment that only bit-flip
errors occur in the network, with a probability $\varepsilon_b$; phase errors
will be added in the next section. The first step of the procedure is to generate
an entangled pair for all the edges of the lattice and to create a
local state $\ket{\Phi^+}$ at the station $O$, see Fig.~\ref{fig:model}. We then
use the entangled pairs to teleport the two qubits of $\ket{\Phi^+}$ to the right
and to the top until they reach $C$, in the following way:
\myItem inner stations receive two qubits from their bottom and left neighbors
and they teleport them to the top and to the right, respectively,
\myItem stations on a boundary do the same, but they either receive one
qubit and ``duplicate'' it (by applying a CNOT with an
additional qubit in the state $\ket{0}$ as target), or they
receive two qubits and ``erase'' one of them (by measuring it in
the $X$ basis and by communicating the outcome result; station $C$ erases
the two qubits it receives),
\myItem the destination stations $A$ and $B$ save each an additional duplicated
qubit; this will create the long-distance entangled state $\ket{\Phi^+}_{AB}$.
%
\subsection{Bit-flip error correction}
\label{ssec:flip-correct}
A syndrome detection and a network-based error correction are now used to
suppress the bit-flip errors that occured while teleporting the initial state from
$O$ to $C$. A very similar procedure can be found in the context of fault-tolerant
error recovery in planar codes \cite{DKLP02}, see discussion in App.~\ref{app:comparison}.\\
%
\paragraph*{Parity checks}
Each station coherently extracts the parity information of the two qubits it receives,
as shown in Fig.~\ref{fig:resources}, and outputs $+1$ if the qubits have the same parity
and $-1$ otherwise (error syndrome).
Stations that receive only one qubit define the parity check output as $+1$.
Given the syndrome pattern for the lattice, the task is now to determine which
are the edges responsible for the bit-flip errors.\\
\begin{figure}
    \begin{center}
        \includegraphics[width=0.96\linewidth]{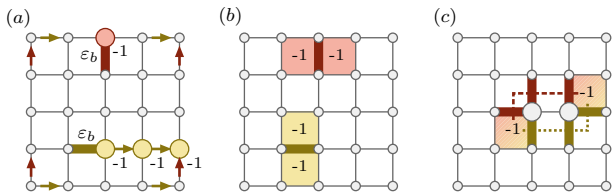}
        \caption{Bit-flip error correction:
            (a) Parity check pattern associated with two erroneous edges, and
            (b) its plaquette representation.
            (c) Two of the three possible edge identifications (with the minimal number
            of errors) for the given syndrome pattern. If the inferred
            edges do not correspond to the erroneous ones, a loop is created and sites lying
            inside this region will apply the wrong bit-flip correction.}
        \label{fig:plaquette}
    \end{center}
\end{figure}
%
\paragraph*{Plaquette representation}
We assign for each plaquette (i.e. for each vertex of the dual lattice) a value
that is the product of the four parity check outputs at its corners.
For example, the parity check pattern and its plaquette representation
induced by one erroneous edge are shown in Fig.~\ref{fig:plaquette}(a,b).
Thus, single erroneous edge results in two $-1$ plaquettes
adjacent to the erroneous edge (except when that edge is on the boundary,
resulting in a single $-1$ plaquette). Erroneous edges can be identified unequivocally
as long as they are \emph{isolated}, but having two (or more) such edges adjacent to the
same plaquette leads to some ambiguity, see Fig.~\ref{fig:plaquette}(c),
and this may cause a bit-flip error in the final state. 
\\
%
\paragraph*{Fidelity of the resulting state}
\begin{figure}
    \begin{center}
        \includegraphics[width=0.95\linewidth]{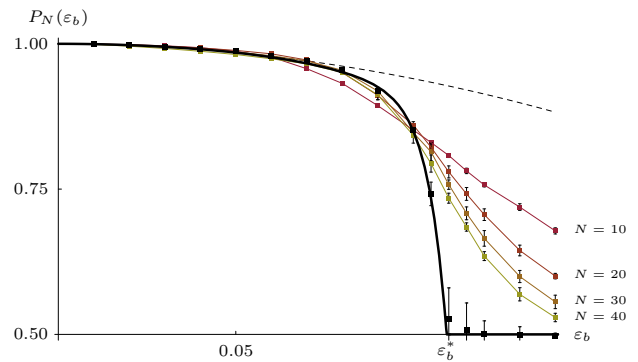}
        \caption{Simulation of the (classical) bit-flip error correction in a
			square lattice with periodic boundary conditions, so that
			syndrom plaquettes always appear in pairs. An unknown bit
			is sent through the lattice and a possible minimal set of erroneous edges
			leading to the corresponding parity check pattern is found using
			Edmonds' algorithm for minimal weight perfect matching \cite{E65}.
			We show here the probability that two random sites infer the
			same value of the bit in a $N \times N$ lattice with bit-flip error probability $\varepsilon_b$.
			The extrapolated function $P_{\infty}$ is plotted in bold line and
			the dashed line represents its behaviour for small $\varepsilon_b$:
			$P_{\infty}(\varepsilon_b)\approx 1-6\,\varepsilon_b^2+\mathcal{O}(\varepsilon_b^3)$.}
        \label{fig:fidelity}
    \end{center}
\end{figure}
Since we assume bit-flip errors only, the resulting state is a mixture of
$\ket{\Phi^+}=\ket{00}+\ket{11}$ and $\ket{\Psi^+}=\ket{01}+\ket{10}$:
\begin{equation}
	\rho_{AB} = F \ketbra{\Phi^+}{\Phi^+} + (1-F) \ketbra{\Psi^+}{\Psi^+}.
\end{equation}
This mixed state is always entangled and can be distilled at a rate of $E=1-H_2(F)$,
which is called distillable entanglement \cite{BBP+96}. Its fidelity $F$, i.e. the probability to
apply the same bit-flip correction at $A$ and $B$, has been calculated numerically,
see Fig.~\ref{fig:fidelity}: this shows that there exists a critical value $\varepsilon_b^*\approx0.11$ above which
the error correction fails to work. A precise value of $\varepsilon_b^*$ is discussed
in App.~\ref{app:comparison}, but we can justify its existence and approximate it
in a simple way by the following argument: each imperfect quantum channel introduces an entropy
$H_2(\varepsilon_b)=-\varepsilon_b\log_2\varepsilon_b-(1-\varepsilon_b)\log_2(1-\varepsilon_b)$
into the network, and each parity check extracts at most 1 bit of information. We thus can maintain an ordered phase if
$2H_2(\varepsilon_b)<1$, i.e. when $\varepsilon_b \lesssim 11\%$.\\
%
\paragraph*{Measurement errors}
In practice the parity check measurements are imperfect and they give a wrong result
with probability $\varepsilon_{c}$. We can generalize the
entropy argument by including the measurement errors, so that at most $1-
H_2(\varepsilon_c)$ bits of information are extracted from a parity check.
The condition for maintaining an ordered phase becomes:
\begin{equation}
    2 H_2(\varepsilon_b) < 2 H_2(\varepsilon_b^*) = 1 - H_2(\varepsilon_c).
\end{equation}
We now try to get rid of the measurement errors by repeating the parity
checks $2r+1$ times and using the majority vote to infer the correct syndrome.
If the parity check measurements do not perturb the qubits, repeating them
can suppress the errors up to $O(\varepsilon_c^{r+1})$. Even if additional
errors are introduced into the system, three repeated measurements can already
help in correcting errors; in fact a measurement error can be treated as an
effective contribution to the error in the channel (see App.~\ref{app:error}),
which approximately becomes $\varepsilon_b'=\varepsilon_b+3\varepsilon_c$,
and an ordered phase is maintained whenever $\varepsilon_b'<\varepsilon_b^*$.\\
%
\paragraph*{Other lattices}
Even if the square lattice is the most natural one we can think of,
ideas of network-based correction can easily be generalized to other geometries.
The triangular lattice, for instance, has a slightly higher critical value:
$\varepsilon_b^*\approx 17\%$. This threshold is found by solving the equation
$3H_2(\varepsilon_b^*)=2$ (similar entropy argument as for the square lattice),
or by considering the corresponding random-bond Ising model on the triangular
lattice (see App.~\ref{app:comparison} and the numerical result in \cite{O07}).\\
%
\paragraph*{Comparison with pure-state percolation}
Our protocol can also be applied when the edges of the lattice are
given by pure but non-maximally entangled states of the form
$\ket{\varphi}=\sqrt{\varphi_0}\ket{00}+\sqrt{\varphi_1}\ket{11}$,
with $\varphi_0\geq\varphi_1$ and $\varphi_0+\varphi_1=1$, and when quantum
operations are supposed perfect. In this situation there already exists a
protocol, based on classical percolation theory, that achieves long-distance
entanglement \cite{ACL07,PCA+08}: one first try to convert each state $\ket{\varphi}$
of the network into the Bell pair $\ket{\Phi^+}$, which is successful with a probability
$p=2\varphi_1$. If $p$ is higher than some threshold $p^*$ depending on the lattice
(for instance $p^*=0.5$ for an infinite square lattice), then there exists with
a strictly positive probability a path of Bell pairs connecting any two sites.
Performing entanglement swappings along such a path finally leads to the
desired long-distance entangled pair. In order to use our approach, we twirl
\cite{VW01} the pure state $\ket{\varphi}$ to the mixed state
\[
	\rho = \frac{(\sqrt{\varphi_0}+\sqrt{\varphi_1})^2}{2}\ketbra{\Phi^+}{\Phi^+} +
		   \frac{(\sqrt{\varphi_0}-\sqrt{\varphi_1})^2}{2}\ketbra{\Phi^-}{\Phi^-},
\]
which corresponds to a bit-flip error probability of $(\sqrt{\varphi_0}-\sqrt{\varphi_1})^2/2$.
Since the threshold for our protocol is about $0.11$, we can deal with states up to
$\varphi_0\lesssim 0.81$, and thus beat the classical percolation protocol which
works only for $\varphi_0\leqslant 0.75$.
%
\subsection{Physical and temporal resources}
\label{ssec:flip-resources}
\begin{figure}
    \begin{center}
        \includegraphics[width=0.96\linewidth]{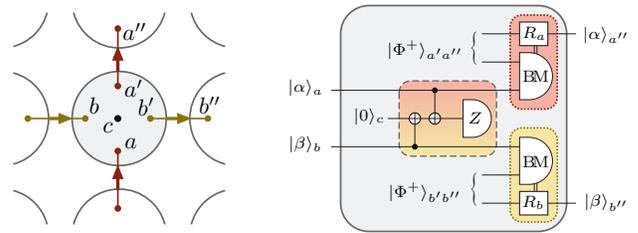}
        \caption{Local resources and operations at each station. At least
            five qubits are needed: one for the parity check measurement
            (dashed box), and four for the teleportations (dotted boxes). The
            choice of the rotations $R$, depending on the outcomes of the Bell
            measurements, can be tracked classically and thus no classical communication
            is necessary between the stations during the process; the interpretation
            of the parity checks can be postponed until its very end.}
        \label{fig:resources}
    \end{center}
\end{figure}

As illustrated in Fig.~\ref{fig:resources}, each station requires:
\myItem four qubits for the connections with its neighbors,
\myItem one qubit for the parity check measurements, which can also be
used to store the final Bell state, and
\myItem about two ancilla qubits (not shown in the figure) to create
the elementary entangled pairs using nested entanglement pumping
(see end of App.~\ref{app:error}).
\myItemEnd
We therefore need approximately seven qubits at each station, and this number
is independent of the size of the lattice.\\
%
\paragraph*{Simultaneous measurements versus quantum memory}
One key advantage of the proposed procedure is that all stations can operate
\emph{simultaneously}, the quantum operations at each station (including
both Bell and parity-check measurements) can be made without knowing the
measurement outcomes from the other stations. The interpretation of these
outcomes, i.e. the choice of the local Pauli frames, can be done once we know
them all. Since the goal is to produce a Bell pair between stations $A$ and $B$,
only the two concerned qubits need to be stored in good quantum memory while waiting
for the measurement outcomes to be collected and analyzed; at the same time
another round of quantum communication in the network can already start.\\
A more favorable application is quantum key distribution. To that purpose,
in fact, all what is needed is the statistical correlation associated with the
Bell pair, rather than the pair itself. This observation allows us to
further relax the requirement of good quantum memory since the ``duplicated'' qubits
at stations $A$ and $B$ can be measured even before the reception of all measurement
outcomes: the two qubits are measured in one of the two complementary bases (e.g. $X$ and $Z$),
which is randomly chosen for each destination station. The outcomes are
secretely stored, while the choice of the basis is publically announced. Once all
measurement outcomes are received, $A$ and $B$ can determine whether they
chose the same basis or not. This is the case with probability one half, and
thus a raw key for cryptography is available. Even if the correlation is not perfect
due to various imperfections, we can apply the procedure of privacy amplification
before we obtain the final highly correlated secrete key, see \cite[p.186]{GRTZ02}.
%
\section{Correcting bit-flip and phase errors}
\label{sec:phase} We now propose a way of suppressing both
bit-flip and phase errors that are present in the quantum network.
%
\subsection{Encoding the qubits}
\label{ssec:phase-encoding}
Each physical qubit considered so far is replaced
by an encoded block of qubits, and we also implement all quantum operations
at the encoded level. Phase errors are suppressed by the code redundancy, and
bit-flip errors are corrected exactly as explained in the previous section.
The subspace for the redundant code of $n=2t+1$ physical qubits is spanned by the
two logical GHZ states:
\begin{equation}\begin{split}
	\ket{\tilde{0}} &\equiv \frac{1}{\sqrt{2}}\left(\ket{+}^{\otimes n}+\ket{-}^{\otimes n}\right),\\
	\ket{\tilde{1}} &\equiv \frac{1}{\sqrt{2}}\left(\ket{+}^{\otimes n}-\ket{-}^{\otimes n}\right).
\end{split}\end{equation}
This choice of code can correct by majority vote up to $t$ phase errors in the block of $n$ qubits,
but it cannot correct any bit-flip errors, which we denote as $(t_p,\,t_b)=(t,0)$.
This is a CSS code with stabilizers generated by $\{X_1X_2,\,X_2X_3,\ldots,\,X_{n-1}X_n\}$,
and all the nice properties of a CSS code can be used, as transversal CNOT gates
or efficient measurements \cite{G98}. Furthermore, all the quantum operations discussed
in the previous section can still be applied, with some minor changes:
\myItem physical qubits are replaced by encoded qubits; in particular we use the
encoded Bell pair $\ket{\tilde{0}\tilde{0}}+\ket{\tilde{1}\tilde{1}}$ as elementary links,
\myItem the CNOT gate is implemented by a transversal CNOT gate between two encoded qubits,
\myItem encoded Pauli operators $\tilde{X}$ and $\tilde{Z}$ are inferred by measuring
all $X$ and $Z$ operators on the qubits of the encoding block, and
\myItem classical error correction is performed to suppress up to $t$ phase errors.
\myItemEnd
It is important for the encoding process to fulfill the requirement of
\emph{fault-tolerance}: the probability to get errors on $j$
physical qubits should be of the order of $\varepsilon_p^j$ for all $j\leq t$
(with $\varepsilon_p$ the phase error probability for a single qubit). Efficient procedures
to fault-tolerantly prepare GHZ states are available, see \cite{K05}
and \cite[Sect.~IX]{JTSL07}. Hence we may treat errors on physical qubits
as independent.
%
\subsection{Required resources for a $N\times N$ lattice}
\label{ssec:phase-sumup}
In order to estimate the fidelity of the long-distance entangled state that is
created using our protocol, one has to quantify the amount of errors that occur at
three different levels: let $\varepsilon\ll1$ be the error probability at the physical
level, $\tilde{\varepsilon}$ at the logical level and $\epsilon$ at the network level.
In App.~\ref{app:error} we find the approximation $\varepsilon\equiv
\varepsilon_b\approx\varepsilon_p\lesssim 8.5\,\beta$,
where $\beta$ is the maximum error probability associated with the local two-qubit
quantum gates, the measurements and the quantum memory. At the encoded level we have to
distinguish the two types of errors, since the encoding preferentially suppresses the phase
errors while it moderately increases the bit-flip errors:
\begin{equation}\begin{split}
    \tilde{\varepsilon}_p &= \sum_{j=t+1}^{2t+1}\begin{pmatrix}2t+1\\j\end{pmatrix}\varepsilon^j
    	\approx\begin{pmatrix}2t+1\\t+1\end{pmatrix}\varepsilon^{t+1},\\
	\tilde{\varepsilon}_b &= \sum_{j=0}^t \begin{pmatrix}2t+1\\2j+1\end{pmatrix}\varepsilon^{2j+1}
		\approx(2t+1)\,\varepsilon.
    \label{eqn:encodedLevelErr}
\end{split}\end{equation}
Phase errors accumulate as the $2N(N-1)$ links of the network are consumed for the teleportations,
and the right bit-flip correction is applied with a probability $P_N(\tilde{\varepsilon}_b)$,
see Fig.~\ref{fig:fidelity}:
\begin{equation}\begin{split}
	\epsilon_p &= 1-(1-\tilde{\varepsilon}_p)^{2N(N-1)}
		\approx 2N(N-1)\,\tilde{\varepsilon}_p,\\
	\epsilon_b &= 1-P_N(\tilde{\varepsilon}_b)
		\approx 1-P_{\infty}(\tilde{\varepsilon}_b)\quad\text{for }N\gg1.
	\label{eqn:networkLevelErr}
\end{split}\end{equation}
This finally results in the state
\begin{multline}
	\rho_{AB} = (1-\epsilon_b)(1-\epsilon_p)\ketbra{\Phi^+}{\Phi^+}
				+ \epsilon_b(1-\epsilon_p)\ketbra{\Psi^+}{\Psi^+}\\
				+ \epsilon_p(1-\epsilon_b)\ketbra{\Phi^-}{\Phi^-}
				+ \epsilon_b\,\epsilon_p\ketbra{\Psi^-}{\Psi^-}.
\end{multline}
By fixing, for example, the distillable entanglement $E=1-H_2(\epsilon_b)
-H_2(\epsilon_p)$ of the final state, and under the conditions that $t$ is an integer and that
$\epsilon_b$ and $\epsilon_p$ should be of the same order of magnitude,
one can now estimate the required resources $t$ (number of qubits used for the encoding)
and $\varepsilon$ (tolerable error probability at the physical level) by solving
Eqs.~(\ref{eqn:encodedLevelErr}, \ref{eqn:networkLevelErr}), see Tab.~\ref{tab:estimates}.
The number $n=2t+1$ scales only
logarithmically with the size of the lattice, and even though $\varepsilon$ decreases with $N$,
it stays of the order of the percent for any realistic quantum network.
\begin{table}
    \begin{center}
    \begin{tabular}{c@{\quad}|@{\quad}c@{\qquad}c@{\quad}c@{\quad}c@{\quad}c@{\quad}c}
        \hline\hline
        $E$		& $N$		& $10^1$	& $10^2$ 	& $10^3$ 	& $10^4$	& $10^5$\\
        \hline
         & $t$				& $2$   	& $3$    	& $4$    	& $5$		& $6$\\[-1ex]
        \raisebox{1.5ex}{0.75}
		 & $\varepsilon$   	& $1.38$	& $0.96$  	& $0.75$  	& $0.62$	& $0.54$\\
		\hline
         & $t$				& $2$   	& $3$    	& $4$    	& $5$		& $6$\\[-1ex]
        \raisebox{1.5ex}{0.50}
		 & $\varepsilon$   	& $1.84$	& $1.27$  	& $0.97$  	& $0.79$	& $0.67$\\
		\hline
         & $t$				& $2$   	& $3$    	& $4$    	& $5$		& $6$\\[-1ex]
        \raisebox{1.5ex}{0.25}
		 & $\varepsilon$   	& $2.02$	& $1.41$  	& $1.08$  	& $0.88$	& $0.74$\\
        \hline\hline
    \end{tabular}
    \caption{Estimation of the resources that are required to create a long-distance pair of
		distillable entanglement $E$ in a $N\times N$ square lattice.
		The size of the encoding is $n=2t+1$ and $\varepsilon$
		is the elementary error probability (in percent) that can be tolerated.
		The total number of qubits at each station is approximatively $5n$.}
    \label{tab:estimates}
    \end{center}
\end{table}
%
\subsection{Universal computation on a line}
\label{ssec:flip-line}
\begin{figure}
    \begin{center}
        \includegraphics[width=0.84\linewidth]{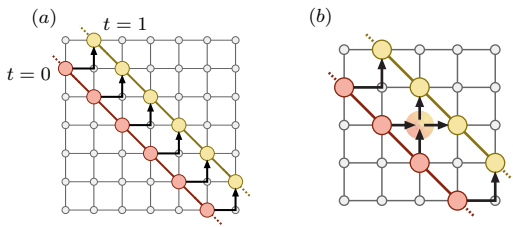}
        \caption{One-dimensional quantum computer embedded in a lattice:
		(a) two teleportations transport the quantum computer between
        time $t$ and $t+1$, and (b) two-qubit gates can be applied
		on neighboring qubits.}
        \label{fig:online}
    \end{center}
\end{figure}

One more general question is whether entanglement distribution in
a two-dimensional lattice, with a fixed local dimension and containing
any kind of errors, is possible at all. Here we show that this question
is related to the existence of fault-tolerant quantum computation in
a one-dimensional setting restricted to next-neighbor gates only.\\
Let us take one diagonal line of the lattice as a one-dimensional
quantum computer at time $t=0$, see Fig.~\ref{fig:online}a. We
can move this quantum computer in the upper-right direction by
teleporting all its qubits to the right and then to the top. The
quantum computer is now supposed to be at time $t=1$ and the errors
that occured during the teleportations are seen as memory errors
between times $t=0$ and $t=1$; we further can transport the
line to reach any time $t$. Any two-qubit gate between neighboring
qubits $a$ and $b$ can be implemented by slightly changing the path of
the teleportation (Fig.~\ref{fig:online}b): one of the qubit is teleported as usual while the other
is first teleported to the top and then to the right such that
both meet at the center station, where the gate is applied.
What we get is a next-neigbor one-dimensional
quantum computation scheme with a simple error model, namely bit-flip
and phase errors occuring randomly with probabilities $\varepsilon_b$
and $\varepsilon_p$.\\
The ``space-like'' task of teleporting a qubit
in a two-dimensional lattice has changed into the ``time-like'' one of
preserving a qubit in a one-dimensional quantum computer. Since
there exist fault-tolerant next-neighbor one-dimensional quantum
computation schemes using two qubits per site \cite{SFH08},
quantum information can be transported over arbitrary distances
if we replace all single-qubit links in our lattice by two-qubit links.
\section{Conclusion}
\label{sec:conclu}
We have proposed a novel way of looking at the problem of entangling two distant
stations in a noisy quantum network. We overcome the main drawbacks of the usual
one-dimensional repeaters, namely the need for good quantum memories or for
fault-tolerant operations via complicated concatenated quantum codes, by correcting
the quantum errors in two distinct ways: the high connectivity of the network
allows us to gain information on the bit-flip errors and to further suppress, while
a somewhat ``classical'' encoding of the qubits prevents the phase errors to propagate.
For any realistic network size we need only a relatively small number of qubits at each
station ($\lesssim 40$), while the tolerable error probability for the various
quantum operations is of the order of the percent. Furthermore, not much effort has
to be put in the storage of the qubits since the proposed scheme is a one-shot
process involving efficient classical algorithms.

\begin{acknowledgments}
We thank J. Preskill for fruitful comments concerning App.~\ref{app:comparison}.
This work has been supported by the QCCC program of the Elite Network of Bavaria (ENB),
the EU projects COVAQIAL and SCALA and the DFG-Forschergruppe 635.
\end{acknowledgments}

\appendix
\numberwithin{equation}{section}

\section{Error model}
\label{app:error}
The error model we consider in this paper includes three major types of errors
relevant to the implementation of the quantum network:
\myItem the infidelity $1-F_0$ of the elementary entangled pairs $\rho$, with
$F_0 = \bra{\Phi^+}\rho\ket{\Phi^+}$,
\myItem a local two-qubit gate error probability $\beta$ and a local measurement error
probability $\delta$, and
\myItem the memory error $\mu\approx\gamma T_{0}$ for a storage time $T_{0}$, assuming
$T_{0}$ to be the time scale for generating encoded Bell pairs between neighboring stations.
\myItemEnd
We use the depolarizing channel for describing an error on a two-qubit gate
$O_{12}^{\text{ideal}}$, see Ref.~\cite{BDCZ98}:
\begin{equation}
    \rho \mapsto O_{12}[\rho] = (1-\beta) O_{12}^{\text{ideal}}[\rho] + \frac{\beta}{4}\,
    \id_{12}\otimes \tr_{12}[\rho],
\end{equation}
and the following POVM for an imperfect measurement on a single qubit:
\begin{equation}
    \left.\begin{aligned}
    P_{0}^{\delta}  &= (1-\delta) \ketbra{0}{0} + \delta \ketbra{1}{1},\\
    P_{1}^{\delta}  &= (1-\delta) \ketbra{1}{1} + \delta \ketbra{0}{0}.
    \end{aligned}\right.
\end{equation}
Equivalently, we can model an imperfect measurement on one qubit by applying the
effective depolarizing channel
\begin{equation}
    O_1^{2\delta}[\rho] = (1-2\delta)\rho + \delta\,\id_1 \otimes \tr_1[\rho]
    \label{eq:app-error-depolarize}
\end{equation}
followed by a perfect measurement. Even though the error probability for this channel
is $2\delta$, the probability to get a wrong measurement outcome is only $\delta$.
Finally, the result of an imperfect memory is modeled by the same depolarizing channel,
with error probability $\mu$.\\
If the initial fidelity $F_{0}$ is not very close to unity, we may use the idea of (nested) entanglement
pumping \cite{DBCZ99} to efficiently pump the Bell pairs to a higher fidelity.
This is only limited by the imperfections of local operations and the decoherence
of the quantum memory, and can lead to a fidelity $F_0'\approx1-\frac{5}{4}\beta-\frac{3}{4}\mu$, see \cite{JTSL07}.\\
%
\paragraph*{Accumulated error on the physical qubits}
With the condition that all operations are performed fault tolerantly, we can
estimate the total error probability which is accumulated on an individual physical
qubit during the creation of a long-distance entangled pair.
First, the probabilities for bit-flip and phase errors \footnote{We conservatively
count errors associated with Pauli operator $Y$ as both bit-flip and phase errors}
associated with the entanglement purification, the local encoding, the CNOT gate
and the quantum teleportation is approximately given by $4\beta+2\delta+\mu/2$.
Then, $m$ repeated parity check measurements may introduce bit-flip and phase
errors with probability $m\,\beta/2$. For $m=3$ rounds, the effective measurement
error probability is about $\beta^2/2+3(\beta+\delta)^2$. Since \emph{one}
measurement error is equivalent to \emph{two} bit-flip errors in two connected
edges, the measurement error of order $(\beta+\delta)^2$ can
be conservatively counted as a bit-flip error of order $\beta+\delta$.
Finally, if we assume $\beta\approx\delta\approx\mu$, we find that the
accumulated probabilities for bit-flip and phase errors are
$\varepsilon\equiv\varepsilon_b\approx\varepsilon_p\lesssim 8.5\,\beta.$

\section{Comparison with the error recovery in planar codes}
\label{app:comparison}
The identification of erroneous edges in our model of quantum networks is very similar
to the error correction for the planar code \cite[Sect.~IV]{DKLP02}, since in both cases we
are looking for a minimal set of edges that match a syndrome pattern. In fact,
since we consider the regime of small bit-flip error probability, the most probable
configurations that lead to a given pattern of plaquettes are the ones that contain
the least number of errors. Plaquettes with an error syndrome appearing in pairs,
the most important part of the error correction is to find a grouping of these ``defects''
that minimizes the sum of the distances within the pairs. Two paired defects have
then to be connected by a path of minimal length, which is in general not unique.
Since all such paths are equally probable, the choice of the right one is ambiguous and loops
of wrongly inferred edges may be created. If the error probability $\varepsilon_b$ is too large,
these loops proliferate and eventually give rise to a chain that stretches from
one boundary to another, suppressing any long-distance correlation.
In an infinite square lattice this happens if $\varepsilon_b$ is larger than the critical value
\begin{equation}
    \varepsilon_b^* \approx 0.1094,
\end{equation}
which has been numerically calculated via a mapping to the two-dimensional random-bond
Ising model \cite{HPP01}.\\
At this point a basic difference between the two models has to be pointed out: in the planar
code, homologically trivial loops do not affect the state used as quantum memory, while
non-contractible ones induce a logical error. In our quantum network, however,
a contractible loop also affects the final state if one of the destination stations
lies in its inside, since the wrong bit-flip error correction is applied to it.
In that sense, the destination stations in our model can be viewed as punched holes
in the planar code, which are used to encode logical qubits.\\
This observation removes part of the ambiguity concerning the choice of the path that connects
two defects: it has to follow (as well as possible) a straight line.
For the syndrome pattern shown in Fig.~\ref{fig:plaquette}(c), for instance, one can easily
verify that the path ``$\rightarrow\uparrow\rightarrow$'' (starting from the bottom-left
plaquette) minimizes the average number $\bar{n}$ of sites that infer a wrong
error correction:
\begin{subequations}
\begin{gather}
	\bar{n}(\rightarrow\uparrow\rightarrow)=2/3,\\
	\intertext{while we have for the two other possible paths:}
	\bar{n}(\uparrow\rightarrow\rightarrow)=\bar{n}(\rightarrow\rightarrow\uparrow)=1.
\end{gather}
\end{subequations}

\bibliography{article}

\end{document}